\documentclass[10pt]{amsart}
\usepackage{a4wide}
\usepackage{amsfonts}
\usepackage{psfrag}
\usepackage{color}
\usepackage[all]{xypic}
\usepackage{hyperref}
\usepackage{amsmath}
\usepackage{amssymb}
\usepackage{amsthm}
\usepackage{graphicx}
\usepackage{mathtools}

\newtheorem{theorem}{Theorem}[section]

\newtheorem{proposition}[theorem]{Proposition}

\theoremstyle{definition}

\newtheorem{remark}[theorem]{Remark}

\newcommand{\PP}{\mathbb{P}}

\newcommand{\E}{\mathrm{e}}
\newcommand{\D}{\mathrm{d}}

\newcommand{\M}{\mathrm{M}}

\newcommand{\Ff}{\mathcal{F}}

\newcommand{\I}{\mathrm{i}}

\newcommand{\Oo}{\mathcal{O}}

\numberwithin{equation}{section}
\numberwithin{equation}{section}
\numberwithin{equation}{section}

\begin{document}
\title{Mass at zero in the uncorrelated SABR model and implied volatility asymptotics}

\author{Archil Gulisashvili}
\address{Department of Mathematics, Ohio University}
\email{gulisash@ohio.edu}

\author{Blanka Horvath}
\address{Department of Mathematics, Imperial College London}
\email{b.horvath@imperial.ac.uk}

\author{Antoine Jacquier}
\address{Department of Mathematics, Imperial College London}
\email{a.jacquier@imperial.ac.uk}
\date{\today}
\thanks{
The authors would like to thank Rama Cont and Josef Teichmann for initiating 
the series of ETH-Imperial College workshops, where this project initiated.
BH would like to thank Leif D\"oring and Leonid Mytnik for stimulating discussions on time change techniques.
BH acknowledges financial support from the SNF Early Postdoc Mobility Grant 165248.
AJ acknowledges financial support from the EPSRC First Grant EP/M008436/1.
The numerical implementations have been carried out on the collaborative platform Zanadu
(\href{www.zanadu.io}{www.zanadu.io}).
}

\maketitle

\begin{abstract}
We study the mass at the origin in the uncorrelated SABR stochastic volatility model,
and derive several tractable expressions, in particular when time becomes small or large.
As an application--in fact the original motivation for this paper--we derive 
small-strike expansions for the implied volatility when the maturity becomes short or large.
These formulae, by definition arbitrage free, allow us to quantify the impact of the mass at zero 
on existing implied volatility approximations, and in particular how correct/erroneous these approximations become.
\end{abstract}

\section{Introduction}
 
The stochastic alpha, beta, rho (SABR) model introduced by Hagan, Kumar, Lesniewski and 
Woodward in~\cite{ManagingSmileRisk, HLW} 
is now a key ingredient--and has become an industry standard--on 
interest rates markets~\cite{AndreasenHuge, AntonovFree, BallandTran, Rebonato}. 
It is defined by the pair of coupled stochastic differential equations

\begin{equation}\label{eq:SABRSDE}
\begin{array}{rlrl}
\D X_t & = Y_t X_t^{\beta}\D W_t, \qquad & X_0 & = x_0> 0,\\
\D Y_t & = \nu Y_t \D Z_t, \qquad & Y_0 & = y_0>0,\\
\D \langle Z,W\rangle_t & = \rho \D t,
\end{array}
\end{equation}
where $\nu>0$, $\rho\in (-1,1)$, $\beta \in (0,1)$,
and $W$ and $Z$ are two correlated Brownian motions on a filtered probability space
$(\Omega, \Ff, (\Ff_t)_{t\geq 0}, \PP)$.
Its popularity arose from a tractable asymptotic expansion of the implied volatility
(derived in~\cite{ManagingSmileRisk}),
and from its ability to capture the observed volatility smile; 
calibration therefore being made easier using the aforementioned expansion.
In today's low interest rate and high volatility environment, 
the implied volatility obtained by this very expansion can however yield 
a negative density function for the price process~$X$ in~\eqref{eq:SABRSDE},
therefore exhibiting arbitrage.

This problem of negative density in low interest-rate environments has been directly addressed 
by Hagan et. al~\cite{HaganArbFreeSABR}, Balland and Tran~\cite{BallandTran},
and Andreasen and Huge~\cite{AndreasenHuge}, 
who proposed modifications of the original SABR model.
There exist several refinements to the asymptotic formula itself:
in~\cite{Obloj} Ob{\l}{\'o}j fine tunes the leading order, 
and Paulot~\cite{Paulot} provides a second-order term. 
In certain parameter regimes the exact density has been derived 
for the absolutely continuous part (on~$(0,\infty)$) of the distribution of~$X$: 
in the uncorrelated case $\rho=0$, formulae were obtained in~\cite{Antonov, Islah} 
by applying time-change techniques.
The correlated case is much harder, and
approximations have been derived using projection methods in~\cite{Antonov, AntonovFree},
and using geometric tools in~\cite{GHJ16}.
Barring computational costs, availability of the distribution of the SABR process is equivalent to computing 
any European prices.
This however calls for a computation, not only on the continuous part of the distribution (on $(0, \infty)$),
but also of its singular part at the origin. 
Absorbing boundary conditions at the origin ensure ensure that the forward rate process~$X$ is a true martingale, 
and the singular part can hence accumulate mass, 
depending on the starting value of the process, the parameter configuration and the time horizon.

The original asymptotic formula typically loses accuracy for long-dated derivatives, 
when the CEV exponent~$\beta$ is close to zero, 
or when the volatility of volatility~$\nu$ is large. 
The parameter~$\beta$ governs the dynamics of the smile, 
and small values thereof are usually chosen when the asymptotic formula fails, namely 
on markets where the forward rate is close to zero 
and for long-dated options~\cite{BallandTran, ManagingSmileRisk}. 
Indeed, it comes as no surprise that the orginal formula--which is an asymptotic expansion for small values 
of~$\nu^2 T$--breaks down for large maturities, 
but it is well known that the reasons for the inconsistencies of the SABR formula are subtler than that.
What we highlight here is that the mass at zero can be held accountable for the irregularities in this case as well.
While standard numerical methods proved reliable when the process remains strictly positive, computing the probability mass of the SABR model at the origin is a more delicate issue. 
Due to the singularity at the origin, usual regularity assumptions ensuring stability of numerical techniques 
(finite differences or Monte Carlo) are violated at this point, 
and a rigorous error analysis for these methods is not (yet) available. 
In addition, producing reference values becomes computationally intensive for short time scales.

Since much of the popularity of the SABR model is due to the tractability of its asymptotic formula, 
one should aim at preserving it while taking into account the mass at zero. 
The parameter sets $\rho=0$ or $\beta=0$ are the most tractable, and in fact 
(as observed in~\cite{DoeringHorvathTeichmann}) the only ones where certain advantageous regularity properties of the SABR process can be expected.
We therefore concentrate here on the singular part of the distribution for these regimes,
that is, we study the probability $\PP(X_T=0)$ and provide tractable formulae and asymptotic approximations.
The relevance of these parameter configurations is emphasized by recent results~\cite{AntonovMixing}, which suggest a so-called `mixture' SABR (a combination of the $\rho=0$ and $\beta=0$ cases) approach to handle negative interest rates in an arbitrage-free way.
From a modelling perspective, one may question the relevance of an absorbing boundary condition at zero 
in a financial context, where negative rates can actually occur.
In fact, from a stochastic analysis perspective, when $\beta=0$ there is no need to impose such a boundary condition. 
Remarkably however, as pointed out in~\cite{AntonovMixing}, even in market conditions where interest rates become negative, the historical evolution of interest rates 
suggests
that their dynamics follow processes whose probability distribution exhibit a singularity at the origin\footnote{That is, rates `stick' to zero for certain periods of time, see~\cite{AntonovMixing} for more details.},
which makes the computation of the mass at zero rates relevant for these market scenarios as well.

A further application is a direct approximation of the left wing of the implied volatility smile.
In order to understand the small-strike behaviour of the SABR smile it is essential to determine the probability mass at the origin:
asymptotic approximations of the implied volatility are available, 
not only for small and large maturities, but also for extreme strikes. 
Roger Lee's celebrated Moment Formula~\cite{RogerLee}--subsequently refined by Benaim and Friz~\cite{BenaimFriz} and 
Gulisashvili~\cite{GulisashviliAsympt10}--relates the behaviour of the implied volatility~$I_T(K)$ for small strike~$K$ and maturity~$T$
to the behaviour of the price process ~$X$ around the origin.
De Marco, Hillairet  and Jacquier~\cite{DMHJ}, and later Gulisashvili~\cite{GulisashviliMass},
showed that when the underlying distribution has an atom at zero, 
the small-strike behaviour of the implied volatility is solely determined by this mass, 
irrespective of the distribution of the process on $(0,\infty)$. 
We shall numerically confirm this in the (uncorrelated) SABR model, 
using approximations of the probability mass, in agreement with~\cite{AntonovWings}.

In Section~\ref{sec:MassZero} we derive explicit formulae for the mass at zero $\PP(X_T=0)$ 
in the SABR model for finite time as well as for large times in the uncorrelated case.
Under this assumption, it is possible to decompose the distribution into a CEV component and an independent stochastic time change. 
Such time change techniques have been applied to the SABR model in the uncorrelated case 
in~\cite{Antonov, ChenOsterlee, Islah} 
to determine the exact distribution of the absolutely continuous part of the distribution on $(0,\infty)$. 
Therefore, our formulae complement these by providing the singular part of the distribution
(see~\cite{Hobson, Veraart} for more details about time change techniques in stochastic volatility models).
In Section~\ref{SS:smalltime} and Section~\ref{section:Largetimeuncor}, 
we derive asymptotic expansions for the density of time-changed Brownian 
motion---inspired by the works of 
Borodin and Salminen~\cite{Borodin}, 
Gerhold~\cite{Gerhold} and 
Matsumoto and Yor~\cite{MY2}---which we use to derive the behaviour of the atom at the origin 
for short and large times.
Finally, in Section~\ref{sec:ImpliedVolatility}, we use these results to determine 
the left wing (small strikes) of the SABR implied volatility.
Using the formulae provided in~\cite{DMHJ, GulisashviliMass},
we highlight the fact that some of the widely used expansions exhibit arbitrage in the left wing, 
and propose a way to regularise them in this arbitrageable region.
\vspace{3mm}

\section{Mass at zero in the uncorrelated SABR model}\label{sec:MassZero}
The price process $X$ in~\eqref{eq:SABRSDE} is a martingale~\cite[Remark 2]{Jourdain}. 
If we consider $X$ on the state space $[0,\infty)$, the origin, which can be attained, 
has to be absorbing~\cite[Chapter III, Lemma 3.6]{Jacod}. 
For two functions $f$ and $g$, we shall write $f(z)\sim g(z)$ as~$z$ tends to zero
whenever $\lim\limits_{z\to 0} f(z)/g(z) = 1$.

\subsection{The decomposition formula for the mass}
In the case where the correlation coefficient~$\rho$ is null, the mass at the origin can be computed semi-explicitly. 
Conditioning on the path of the volatility process~$Y$, 
the resulting process~$\widehat{X}$ satisfies the CEV stochastic differential equation
$$
\D\widehat{X}_t = \widehat{Y}_t\widehat{X}_t^{\beta}\D W_t,
$$
starting from $\widehat{X}_0=x_0$,
where $\widehat{Y}$ is a deterministic time-dependent volatility coefficient, and represents, 
for fixed~$\omega\in\Omega$, a realisation of the paths of~$Y$.
Consider now the simple CEV equation
$\D \widetilde{X}_t=\widetilde{X}_t^{\beta}\D W_t$
starting from $x_0$, and set 
$$
\widehat{G}_t := \frac{\widehat{X}_t^{2(1-\beta)}}{(1-\beta)^2}
\qquad\mbox{and}\qquad
\widetilde{G}_t := \frac{\widetilde{X}_t^{2(1-\beta)}}{(1-\beta)^2}.
$$
Then 
$\widehat{G}_t=Z_{\int_0^t\widehat{Y}_s^2\D s}$,
where $Z$ is a Bessel process satisfying the SDE~\cite[Subsection 1.1]{Islah}
$$
\D Z_t=\frac{1-2\beta}{1-\beta}\D t+2\sqrt{|Z_t|}\D W_t,
\qquad Z_0 = \frac{x_0^{2(1-\beta)}}{(1-\beta)^2}.
$$
By It\^o's formula, the process $\widetilde{G}$ solves the same SDE, so that $Z=\widetilde{G}$, and therefore
$\widehat{X}=\widetilde{X}_{\int_{0}^{\cdot} \widehat{Y}_s^2\D s}$.
It follows that $X$ can be obtained from~$\widetilde{X}$ using the stochastic time change
$t\mapsto\int_0^t Y_s^2 \D s$,
namely
$X_t=\widetilde{X}_{\int_0^t Y^2_s \D s}$.
Since this time change is independent of~$\widetilde{X}$,
one can decompose the mass at zero of the SABR model into that of the CEV component at zero 
and the density of the time change:
\begin{equation}\label{eq:MassAtZeroCalc}
\PP\left(X_{t}=0\right)
 = \int_0^{\infty} \PP \left(\widetilde{X}_{r}=0\right) \PP\left(\int_0^t Y_s^2 \D s\in \D r\right)\D r,
\end{equation}
where the mass at zero in the CEV model is given by 
(see~\cite{DMHJ} or~\cite[Section 6.4.1]{jeanblanc2009mathematical})
\begin{equation}\label{eq:MassZeroCEV}
\PP\left(\widetilde{X}_{r}=0\right)
 = 1-\Gamma \left(\frac{1}{2(1-\beta)},\frac{x_0^{2(1-\beta)}}{2r(\beta-1)^2}\right),
\end{equation}
with $\Gamma$, the normalised lower incomplete Gamma function:
$\Gamma(v,z) \equiv \Gamma(v)^{-1}\int_0^zu^{v-1}\E^{-u} \D u$.
\begin{remark}
If $\beta \in [1/2,1)$ in~\eqref{eq:SABRSDE}, the origin is naturally absorbing, 
and the mass at zero is given by~\eqref{eq:MassZeroCEV}. 
When $\beta\in[0,1/2)$, the solution to ~\eqref{eq:SABRSDE} is not unique, 
and a boundary condition at the origin has to be imposed.  
Should one consider the origin to be reflecting, the transition density would then become norm preserving,
and no mass at the origin would be present. However, it is easy to see that there is an arbitrage opportunity if the origin is reflecting. 
Formula~\eqref{eq:MassZeroCEV} carries over to the case $\beta\in[0,1/2)$ when the origin is assumed to be absorbing, which we shall always consider from now on.
This is of course in line with~\cite[Chapter III, Lemma 3.6]{Jacod}, mentioned above,
which states that the origin has to be absorbing for a non-negative supermartingale.
\end{remark}
Since for each $s\geq 0$, $Y_s$ is lognormally distributed, we can write
\begin{equation}\label{DensityYToZ}
\PP\left(\int_0^t Y_s^2  \D s\in \D r\right)
=\PP\left(\int_0^{t} \exp\left(2\nu Z^{(-\nu/2)}_s\right) \D s \in \D \widetilde{r}\right),
\end{equation}
where $\widetilde{r}:=\frac{r}{y_0^2}$, $Z^{(-\nu/2)}_s:= Z_s -\frac{1}{2}\nu s$;
the density of this functional is given by~\cite[Formula 1.10.4]{Borodin}
\begin{equation}\label{eq:BorodinProba}
\PP\left(\int_0^{t} \E^{2\nu Z^{(-\nu/2)}_s} \D s \in \D \widetilde{r}\right)
 = \frac{2^{1/4}\sqrt{\nu}}{\widetilde{r}^{3/4}}
\exp\left(-\frac{\nu^2 t}{8}-\frac{1}{4\nu^2 \widetilde{r}}\right)
m_{2 \nu^2 t}
\left(-\frac{3}{4}, \frac{1}{4 \nu^2 \widetilde{r}}\right)\D \widetilde{r},
\end{equation}
where the function~$m$ is defined as~\cite[page 645]{Borodin}:
\begin{equation}\label{eq:m}
m_y(\mu,z) \equiv 
\frac{8 z^{3/2} \Gamma(\mu+\frac{3}{2})\E^{\frac{\pi ^2}{4y}}}{\pi \sqrt{2 \pi y}} 
\int_0^{\infty} \E^{-z \cosh (2u) - \frac{1}{y}u^2} \M\left(-\mu,\frac{3}{2}, 2 z \sinh (u)^2\right) 
\sinh(2u)\sin\left(\frac{\pi u}{y}\right) \D u,
\end{equation}
and where the Kummer function~$\M$ reads
\begin{equation}\label{Kummer}
\M(a,b,x) \equiv 1+ \sum_{k=1}^{\infty} \frac{a(a+1)\ldots(a+k-1)x^k}{b(b+1)\ldots(b+k-1)k!}.
\end{equation}

\subsection{Small-time asymptotics}\label{SS:smalltime}
We now study the behaviour of the mass at zero 
$\PP\left(X_{t}=0\right)$ as time becomes small.
The main challenge is to provide a short-time asymptotic formula for the density of the time change process, for which standard expansion techniques are not applicable. 
The additive functional arising from the density of an integral over the exponential of Brownian motion often appears in the pricing of Asian options and is of interest on its own. 
This density is notoriously difficult to evaluate in small time, 
due to a highly oscillating factor connected to the Hartman-Watson distribution~\cite{MY, MY2} and \cite[Section 4.6]{Gulisashvili}. 
These numerical issues are discussed in~\cite{Barrieu},
and Gerhold~\cite{Gerhold} used saddlepoint methods to provide short-time estimates.
Because of the time change and the complexity of the Kummer function (in the integrand), 
small-time asymptotics of the mass at zero cannot be estimated directly. 
Instead, we use an inverse Laplace transform approach, inspired by~\cite{Gerhold},
to provide small-time asymptotic estimates for the density of the time change. 
From~\eqref{eq:BorodinProba} and~\eqref{eq:m}, we introduce the notation $y:=2\nu^2 t$, 
and we shall alternate between the two notations without ambiguity in order to simplify some of the formulations below.
\begin{remark}
For $\varpi:=1/y$, the function~$m$ has the form 
$m_\varpi(\cdot) = c_\varpi\int_{0}^{\infty}\E^{-u^2\varpi} f_\varpi(u) \D u$,
for some $c_{\varpi}$ and $f_{\varpi}$. 
One might be tempted to use a standard Laplace method to determine the behaviour of~$m_\varpi$ 
as~$\varpi$ tends to infinity.
However, at the saddlepoint $u^*=0$, attained at the left boundary of the integration domain, all the derivatives of the function $f_\varpi$--appearing as coefficients of the expansion--are null,
and the method does not apply.
\end{remark}
We now formulate one of the main results of the paper, 
which characterises the small-time behaviour of the mass at zero in the uncorrelated SABR model. 
For every $r, y > 0$, let~$u_y$ denote the largest (positive) solution to the equation
\begin{equation}\label{eq:Saddlepoint}
2\mu - 1 + 4 u y + 2\log(z/2)\sqrt{u} -\sqrt{u}\log(u) = 0,
\end{equation}
with $z := \frac{y_0^2}{4\nu^2r}$. 
Clearly, $u_y$ depends on $r$, but we shall omit this dependence in the notation. 
Set
\begin{equation}\label{eq:My}
M_y := \frac{\log(u_y)}{16 u_y^{3/2}} - \frac{\alpha }{8 u_y^{3/2}} + \frac{1-2\mu}{8u_y^2}.
\end{equation}
The following theorem, based on~\eqref{eq:MassAtZeroCalc} and~\eqref{eq:MassZeroCEV},
provides a short-time estimate for the mass at zero.
As showed in the proof, the expansion of the integrand is performed using saddlepoint analysis 
and complex contour deformation.
Precise error estimates however require substantial additional work and new techniques, 
which we hope to develop in future publication.
The numerics performed later in the paper strongly confirm our result.
\begin{theorem}\label{thm:expansionP}
In the uncorrelated SABR model, the asymptotic equivalence 
$$\PP\left(X_{t}=0\right)
 \sim 
\frac{y_0^{3/2}\E^{5/4}}{2^{7/4}\sqrt{\nu\pi}}\exp\left(-\frac{\nu^2 t}{8}\right)
\int_0^{\infty}\exp\left\{\frac{\log(u_y)}{2}\left(\mu - \frac{1}{2}\right) - u_y y + \sqrt{u_y}\right\} \frac{g(r)}{\sqrt{M_y}}
\D r,
$$
holds as~$t$ tends to zero, where 
$\displaystyle
g(r)\equiv \PP \left(\widetilde{X}_{r}=0\right)\frac{1}{r^{5/4}}\exp\left(-\frac{y_0^2}{4\nu^2 r}\right)
$.
\end{theorem}
Theorem~\ref{thm:expansionP} follows from~\eqref{eq:MassAtZeroCalc} and the following assertion.
\begin{proposition}\label{propo:expansionP}
As $y$ (equivalently $t$) tends to zero, we have (recall that $y=2\nu^2 t$)
$$
\PP\left(\int_0^t Y_s^2  \D s\in \D r\right)
\sim \frac{y_0^{3/2}\E^{5/4}}{r^{5/4}2^{7/4}\sqrt{\nu\pi}}
\exp\left(-\frac{y}{16}-\frac{y_0^2}{4\nu^2 r}\right)
\exp\left[\frac{\log(u_y)}{2}\left(\mu - \frac{1}{2}\right) - u_y y + \sqrt{u_y}\right]
\frac{\D r}{\sqrt{M_y}}.
$$
\end{proposition}

The technical part of the proof relies on the following proposition, proved in Appendix~\ref{sec:ProofPropExpansion}.

\begin{proposition}\label{prop:expansionMy}
As $y$ tends to zero, the function~$m_y$ in~\eqref{eq:m} satisfies
$$
m_{y}(\mu,z)
 \sim \frac{\sqrt{z}\exp\left(\frac{1}{2}-\mu\right)}{2\pi}
\exp\left[\frac{\log(u_y)}{2}\left(\mu - \frac{1}{2}\right) - u_y y + \sqrt{u_y}\right]
\sqrt{\frac{\pi}{M_y}}.
$$
\end{proposition}
\begin{remark}
The proof of the proposition uses saddlepoint analysis.
The saddlepoint~$u_y$ is the solution to~\eqref{eq:Saddlepoint},
but does not admit a closed-form expression; 
however, as seen in the proof, it is possible to expand it as $y$ tends to zero to obtain
$$
m_{y}(\mu, z) = \frac{\sqrt{z}|\log(y)|}{2\sqrt{\pi}}
\exp\left\{-\frac{\log(y)^2}{4y} + \frac{|\log(y)|}{2y}
+\left(\frac{1}{2}-\mu\right)\left[1 - \log\left(\frac{|\log(y)|}{2y}\right)\right]\right\} \\
\left[\frac{1}{y^{3/2}}+\mathcal{O}\left(y^{3/2}\right)\right],
$$
but numerical computations however show that this estimate is not very accurate.
\end{remark}

\subsection{Large-time asymptotics}\label{section:Largetimeuncor}
We now concentrate on the large-time behaviour of the mass at zero in the uncorrelated SABR model.
From~\cite[Formula 1.8.4, page 612]{Borodin}, the formula
$$
\PP\left(\int_0^{\infty} \exp\left(2\nu Z^{(-\nu/2)}_s\right) \D s \in \D \widetilde{r}\right)
=\frac{\widetilde{r}^{-3/2}}{\nu\sqrt{2\pi}}\exp\left(-\frac{1}{2 \nu^2 \widetilde{r}}\right)\D \widetilde{r}
$$
holds, so that the decomposition~\eqref{eq:MassAtZeroCalc} together with~\eqref{DensityYToZ}
imply that (recall that $\widetilde{r} = \frac{r}{y_0^2}$)
\begin{align}\label{eq:Mass0LargeTime}
\PP_\infty:=\lim_{t\uparrow\infty}\PP(X_t = 0) 
& = 
\frac{y_0}{\nu\sqrt{2 \pi}}\int_{0}^{\infty}\left[1-\Gamma \left(\frac{1}{2(1-\beta)},\frac{x_0^{2(1-\beta)}}{2r(\beta-1)^2}\right)\right]
r^{-3/2}\exp\left(-\frac{y_0^2}{2 \nu^2 r}\right)\D r\nonumber\\
& = 
1 - 
\frac{y_0}{\nu\sqrt{2 \pi}}\int_{0}^{\infty}\Gamma \left(\frac{1}{2(1-\beta)},\frac{x_0^{2(1-\beta)}}{2r(\beta-1)^2}\right)
r^{-3/2}\exp\left(-\frac{y_0^2}{2 \nu^2 r}\right)\D r.
\end{align}
When $\beta=0 (=\rho)$, 
the SABR model~\eqref{eq:SABRSDE} reduces to a Brownian motion on the hyperbolic plane
(up to a deterministic time change), 
and a simple computation shows that~\eqref{eq:Mass0LargeTime} simplifies to
$$
\left.\PP_\infty\right|_{\beta=0} = 1 - \frac{2}{\pi}\arctan\left(\frac{\nu x_0}{y_0}\right).
$$
When $\beta\ne 0$, the integral in~\eqref{eq:Mass0LargeTime} does not have a closed-form expression.
Expanding the exponential factor for small $y_0$, we can however write,
for any $n\in\mathbb{N}$, the $n$th-order approximation 
\begin{align*}
\PP_\infty^{(n)}
& := \int_{0}^{\infty}\left[1-\Gamma \left(\frac{1}{2(1-\beta)},\frac{x_0^{2(1-\beta)}}{2r(\beta-1)^2}\right)\right]
\frac{y_0}{\nu r^{3/2}\sqrt{2 \pi}}\sum_{k=0}^{n}\frac{1}{k!}\left(-\frac{y_0^2}{2 \nu^2 r}\right)^k\D r\\
& = \sum_{k=0}^{n}\frac{y_0^{2k+1}}{k!\nu \sqrt{2\pi}}\left(-\frac{1}{2 \nu^2}\right)^k 
\int_{0}^{\infty}\left[1-\Gamma \left(\frac{1}{2(1-\beta)},\frac{x_0^{2(1-\beta)}}{2r(\beta-1)^2}\right)\right]
r^{-(k+3/2)}\D r\\
& = \frac{2y_0(1-\beta)}{\Gamma\left(\frac{1}{2(1-\beta)}\right)\nu \sqrt{\pi}x_0^{1-\beta}}
\sum_{k=0}^{n}\frac{(-1)^k}{k!}\left(\frac{y_0^2(\beta-1)^2}{\nu^2x_0^{2(1-\beta)}}\right)^k 
\frac{\Gamma\left(k+1+\frac{\beta}{2-2\beta}\right)}{(1+2k)}.
\end{align*}
Note in particular that 
\begin{equation}\label{eq:PP0Infinity}
\PP_\infty^{(0)} =
\frac{2\Gamma\left(1+\frac{\beta}{2-2\beta}\right)}{\Gamma\left(\frac{1}{2-2\beta}\right)}\frac{y_0(1-\beta)}
{\nu \sqrt{\pi} x_0^{1-\beta}}.
\end{equation}
When $r$ tends to infinity, the integrand clearly converges to zero fast enough.
Using the properties of Gamma functions in~\cite[Chapter 6]{Abramowitz}, 
the asymptotic behaviour
$$
1-\Gamma\left(a, \frac{1}{r}\right) \sim \frac{r^{1-a}\exp(-1/r)}{\Gamma(a)}
$$
holds as $r$ tends to zero, ensuring that the integral is well defined for all $n\in \mathbb{N}$.
Theorem~\ref{thm:SequencePPn} below shows how well (and when) the sequence~$\PP_\infty^{(n)}$ approximates the mass at zero~$\PP_\infty$. 
Using the Taylor formula with Lagrange's form of the remainder, we obtain
$$
\exp\left(-\frac{y_0^2}{2\nu^2r}\right)
 = \sum_{k=0}^n(-1)^k\frac{1}{k!}\left(\frac{y_0^2}{2\nu^2r}\right)^k
 + \frac{(-1)^{n+1}}{(n+1)!}\E^{-\theta}\left(\frac{y_0^2}{2\nu^2r}\right)^{n+1},
$$
for some $\theta \in (0, y_0^2/(2\nu^2r))$.
Therefore,
\begin{equation}
\left| \exp\left\{-\frac{y_0^2}{2\nu^2r}\right\}-\sum_{k=0}^n(-1)^k\frac{1}{k!}
\left(\frac{y_0^2}{2\nu^2r}\right)^k \right|\le
\frac{1}{(n+1)!}\left(\frac{y_0^2}{2\nu^2r}\right)^{n+1}.
\label{E:b1}
\end{equation}
For any $n\geq 0$, set
\begin{equation}\label{E:b2}
b_n := \frac{2y_0(1-\beta)}{\Gamma\left(\frac{1}{2(1-\beta)}\right)\nu \sqrt{\pi}x_0^{1-\beta}}
\left(\frac{y_0^2(\beta-1)^2}{\nu^2x_0^{2(1-\beta)}}\right)^n
\frac{\Gamma\left(n+1+\frac{\beta}{2-2\beta}\right)}{n!(1+2n)},
\end{equation}
so that from~\eqref{E:b1},~\eqref{E:b2}, and the definitions of $\PP_{\infty}$ and $\PP_\infty^{(n)}$,
it follows that, for any $n\geq 0$,
\begin{equation}\label{E:b3}
\PP_\infty^{(n)}=\sum_{k=0}^n(-1)^kb_k
\qquad\text{and}\qquad
\left| \PP_{\infty}-\PP_\infty^{(n)} \right|\le b_{n+1}.
\end{equation}
\begin{theorem}\label{thm:SequencePPn}
The following statements hold for the sequence $\PP^{(n)}$ in~\eqref{E:b3}:
\begin{enumerate}
\item[(i)] if $y_0^2(\beta-1)^2 > \nu^2 x_0^{2(1-\beta)}$, or $y_0^2(\beta-1)^2=\nu^2 x_0^{2(1-\beta)}$ and 
$\frac{2}{3}\le\beta< 1$, then the sequence~$(\PP_\infty^{(n)})_{n\ge 0}$ diverges, 
and hence cannot be an approximation to the mass at zero~$\PP_\infty$;
\item[(ii)] if $y_0^2(\beta-1)^2 < \nu^2x_0^{2(1-\beta)}$, or $y_0^2(\beta-1)^2=\nu^2 x_0^{2(1-\beta)}$ and 
$0\le\beta<\frac{2}{3}$ then
\begin{equation}
\PP_{\infty}
 = \PP_\infty^{(n)} + \mathcal{O}\left(n^{-1+\frac{\beta}{2-2\beta}}
 \exp\left(-n\log\left(\frac{\nu^2x_0^{2(1-\beta)}}{y_0^2(\beta-1)^2}\right)\right)\right),
 \quad\text{as }n\text{ tends to infinity}.
\label{E:addit}
\end{equation}
\end{enumerate}
\end{theorem}
\begin{remark}
Remember that $x_0$ denotes the initial value of the stock price or interest rate.
For all practical and sensible values of the parameters, condition~(ii) in the theorem is always in force.
\end{remark}
\begin{proof}
From~\eqref{E:b2}, Stirling's formula for the Gamma function yields, as $k$ tends to infinity,
\begin{equation}\label{E:theo4}
b_k\sim\frac{y_0(1-\beta)}{\Gamma\left(\frac{1}{2(1-\beta)}\right)\nu\sqrt{\pi}x_0^{1-\beta}}k^{-1+\frac{\beta}{2-2\beta}}
\left(\frac{y_0^2(\beta-1)^2}{\nu^2x_0^{2(1-\beta)}}\right)^k,
\end{equation}
From~\eqref{E:b3}, if the conditions of Theorem~\ref{thm:SequencePPn}(i) hold, 
the general term of the series $\sum_{k=0}^{\infty}(-1)^kb_k$ does not tend to zero, 
and the sequence $\PP_\infty^{(\cdot)}$ diverges.
If the conditions of Theorem~\ref{thm:SequencePPn}(ii) hold, 
then~\eqref{E:b3} and~\eqref{E:theo4} imply (\ref{E:addit}), 
which completes the proof of Theorem~\ref{thm:SequencePPn}.
\end{proof}

For practical purposes, depending on the conditions for convergence 
of the sequence~$(\PP_\infty^{(n)})_{n\geq 0}$ in Theorem~\ref{thm:SequencePPn},
it may or may not be useful to use directly the integral form~\eqref{eq:Mass0LargeTime}.
For $(y_0, \nu, \beta, x_0) = (0.1,1.0,0.2,0.2)$ (for which convergence holds),
the mass at zero in this case is $\PP_\infty = 20.833\%$.
Using Theorem~\ref{thm:SequencePPn}, the table below computes the error 
using the sequence~$(\PP_\infty^{(n)})$:\\

\begin{tabular}{|{c}|{c}{c}{c}{c}{c}{c}|}
\hline
\text{} & $n=0$ & $n=1$ & $n=2$ & $n=3$ & $n=4$ & $n=5$ \\
\hline
$| \PP - \PP_\infty^{(n)}|$ 
& 6.43E-3 & 3.41E-4 & 2.13E-05 & 1.43E-06 & 1.01E-07 & 7.29E-09\\
\hline
Computation time (in seconds) &
6.8E-05 & 8.6E-05 & 1.3E-4 & 1.9E-4 & 2.2E-4 & 2.6E-4\\
\hline
\end{tabular}
\\
\vspace{0.1cm}

and the table below computes the integral~\eqref{eq:Mass0LargeTime}
using the Python scipy toolpack for quadrature;
the integral is truncated at some arbitrary value $R>0$:\\

\begin{tabular}{|{c}|{c}{c}{c}{c}{c}{c}|}
\hline
\text{} & $R=20$ &  $R=40$ &  $R=60$ &  $R=80$ &  $R=100$ &  $R=120$\\
\hline
Absolute error
& 2.33E-4 & 1.07E-4 & 6.77E-05 & 4.90E-05 & 3.81E-05 & 3.11E-05\\
\hline
Computation time (in seconds) &
7.6E-3 & 7.9E-3 &  8.9E-3 &  9.2E-3 &  9.6E-3 &  9.9E-3\\
\hline
\end{tabular}
\vspace*{0.1cm}\\
\\
These results suggest that convergence of the series expansion is extremely fast.
In particular, event the limit~\eqref{eq:PP0Infinity}, with $n=0$, yields a very accurate result, 
which allows for a simple interpretation of the impact of each parameter of the model 
on the large-time mass at the origin.

\begin{remark}\label{rem:MonteCarlo}
One could in principle compare these values with Monte Carlo simulations.
However, as far as we are aware, no rate of convergence for such schemes has yet been proved 
for the SABR model, so one may question numbers generated by simulation. 
In addition, it is known~\cite{ChenOsterlee} that in the critical region around zero, 
Monte Carlo methods are prone to a simulation bias.
Nevertheless, for a comparison with the above results we included some corresponding values for the mass generated by a Monte Carlo algorithm in Section~\ref{sec:LargetimeuncorNum} below.
\end{remark}

\subsubsection{Large-time numerics}\label{sec:LargetimeuncorNum}
\begin{figure}[h!]
\begin{center}
\includegraphics[scale=0.45]{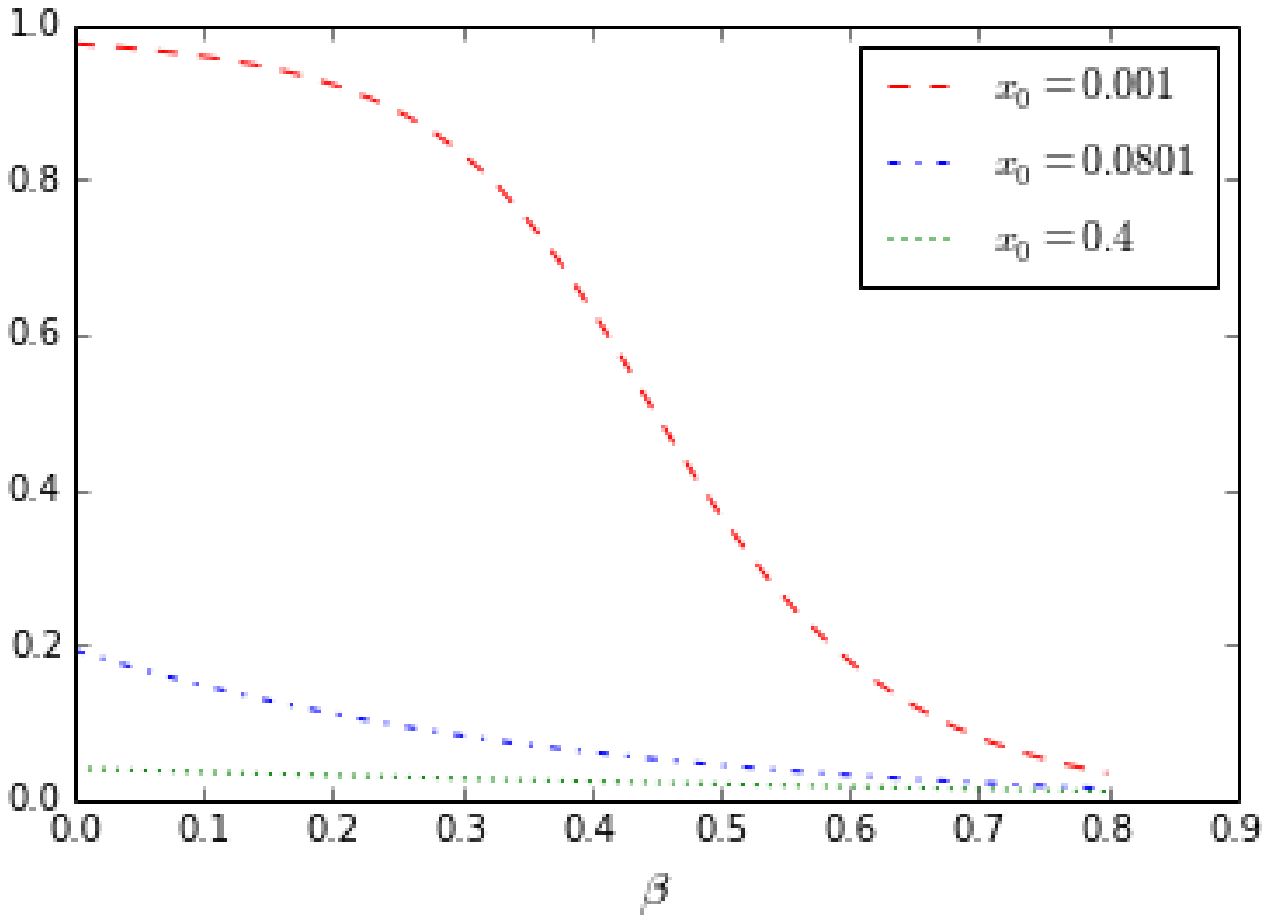}\quad
\includegraphics[scale=0.45]{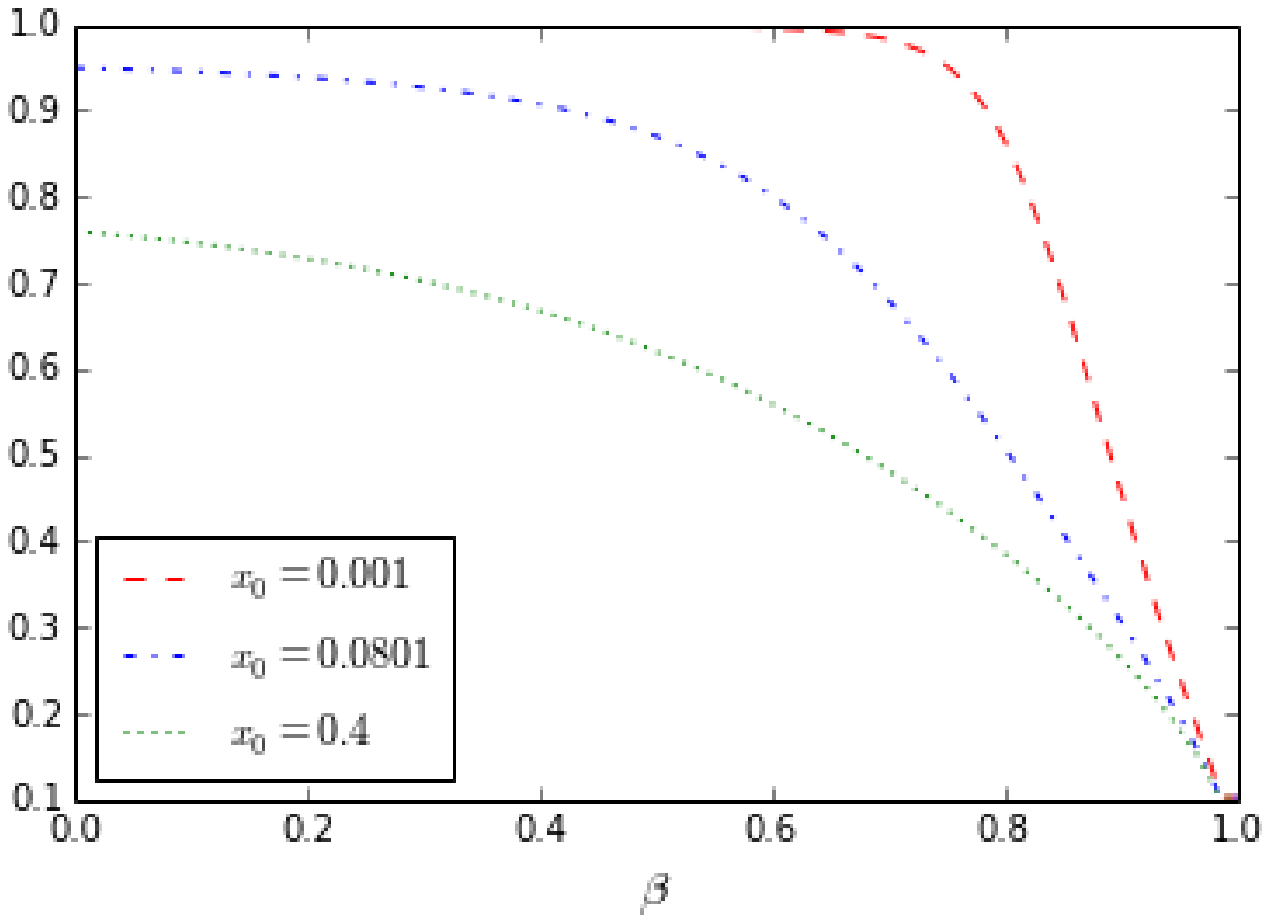}
\caption{Influence of $\beta$ on the large-time mass at zero
in the uncorrelated SABR model with $(y_0, \nu)=(0.015, 0.6)$ (left)
and $(y_0, \nu)=(0.1, 1)$ (right).}
\label{fig:Mass0InfTimeBeta}
\end{center}
\begin{center}
\includegraphics[scale=0.45]{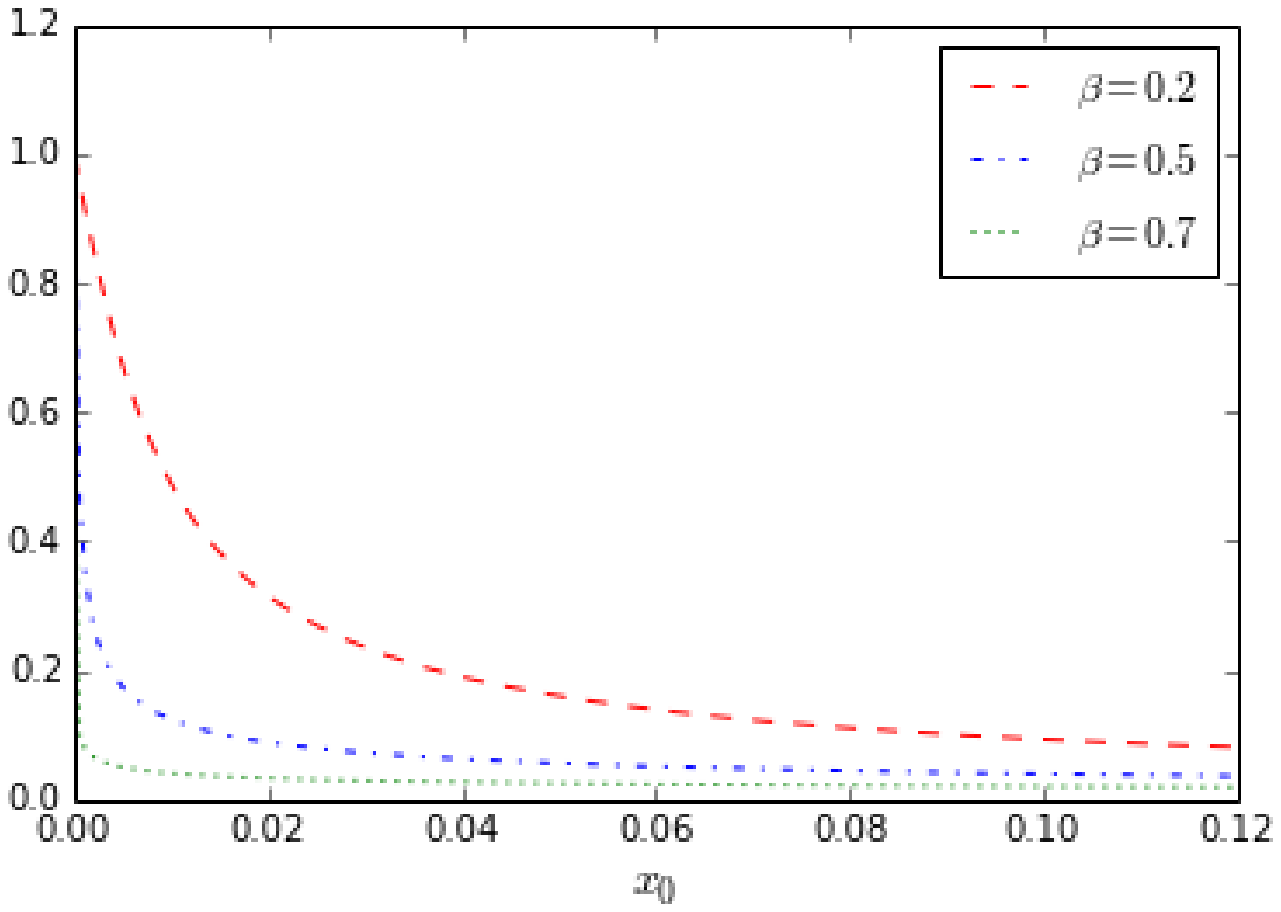}\quad
\includegraphics[scale=0.45]{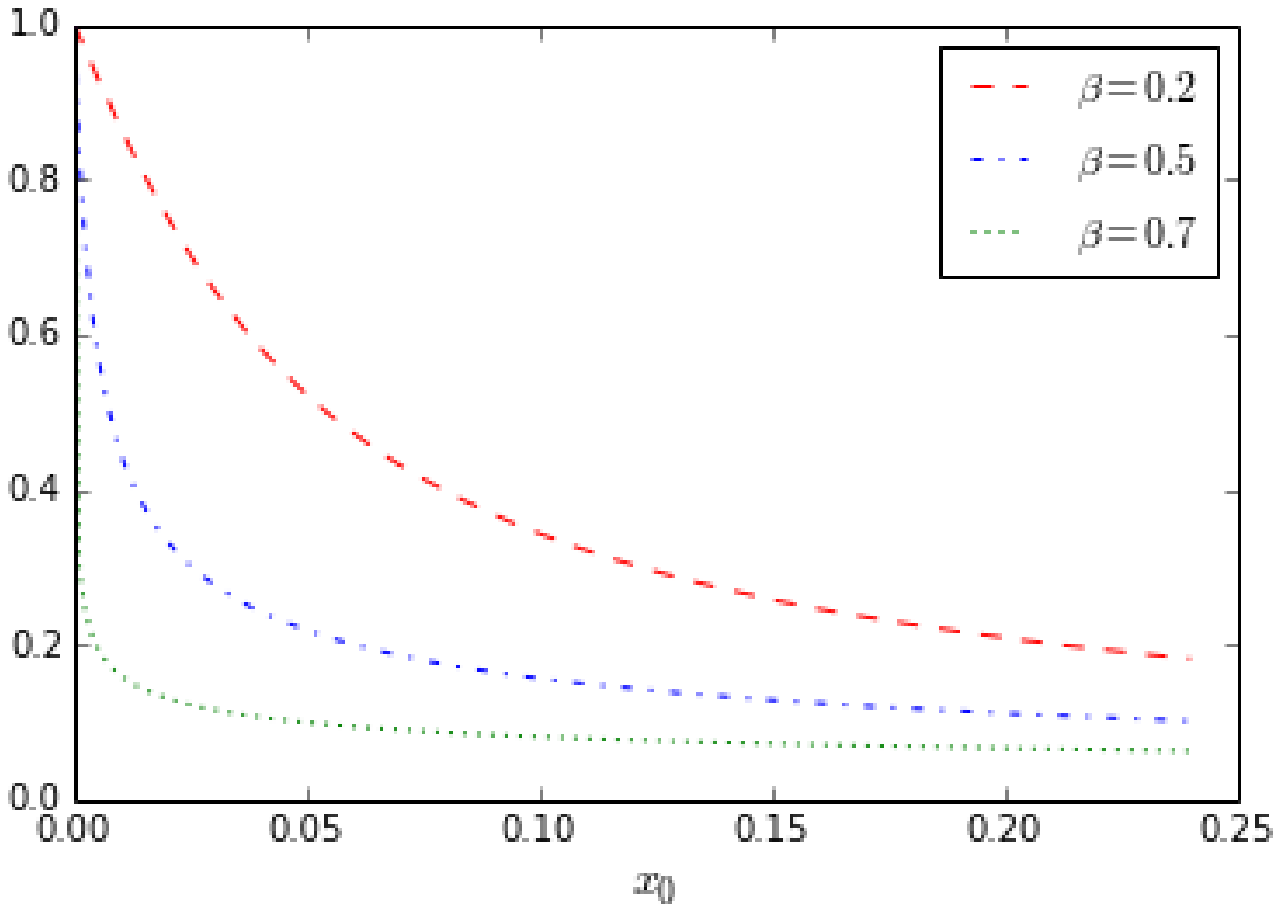}
\caption{Influence of the initial value $x_0$ on the large-time mass at zero
with $(y_0, \nu)=(0.015, 0.6)$ (left) 
and $(y_0, \nu)=(0.1, 1)$ (right).
This gives a numerical interpretation of `feeling the boundary':
as we start the diffusion far enough from the origin, the mass at zero becomes small.}
\label{fig:Mass0InfTimeX0}
\end{center}
\end{figure}

We provide below some numerics of the large-time mass at zero derived in~\eqref{eq:Mass0LargeTime}.
In particular, we observe the influence of the parameter~$\beta$ (Figure~\ref{fig:Mass0InfTimeBeta})
as well as that of the starting point~$x_0$ (Figure~\ref{fig:Mass0InfTimeBeta})
of the uncorrelated SDE~\eqref{eq:SABRSDE}.
As $\beta$ tends to one (from below), the mass at zero is diminishing, 
even for arbitrarily small values of $x_0$. 
Likewise, as the initial value $x_0$ increases, the mass at the origin decreases even for $\beta=0$. 
We shall further comment on the importance of the mass at the origin 
in financial modelling in Section~\ref{sec:ImpliedVolatility} below.

With due caution with respect to their validity (Remark~\ref{rem:MonteCarlo}), 
we include for comparison a sample of values for the mass at zero obtained by Monte Carlo simulations 
with $M=1000$ and $M=2000$ paths, and the corresponding computation times, for different time horizons~$T$. 
The results suggest that the `large-time' regime is already achieved for maturities equal to~$15$ years. 
An explanation for this phenomenon is provided in~\cite[Section 4]{DoeringHorvathTeichmann}.
As in Section~\ref{section:Largetimeuncor} above, we used the parameters 
$(y_0, \nu, \beta, x_0) = (0.1,1.0,0.2,0.2)$, for which the exact mass at zero is $\PP_\infty = 20.833\%$.
\\
\vspace*{0.1cm}\\ 
\begin{tabular}{|{c}|{c}{c}{c}{c}{c}{c}|}
\hline
\text{} & $T=10$ &  $T=15$ &  $T=20$ &  $T=30$ &  $T=50$ &  $T=100$\\
\hline
Monte Carlo mass (M=1000)
& 0.1889 & 0.2020 & 0.2100 & 0.2110 & 0.2050 & 0.2090\\
\hline
Computation time (in seconds) &
3.222 & 3.185 &  3.221 &  3.179 &  3.163 &  3.178\\
\hline
\end{tabular}
\vspace*{0.5cm}\\
\begin{tabular}{|{c}|{c}{c}{c}{c}{c}{c}|}
\hline
\text{} & $T=10$ &  $T=15$ &  $T=20$ &  $T=30$ &  $T=50$ &  $T=100$\\
\hline
Monte Carlo mass (M=2000)
& 0.2100 & 0.2075 & 0.2050 & 0.2100 & 0.2065 & 0.2175\\
\hline
Computation time (in seconds) &
6.4437 & 6.8386 & 6.4805  &  7.6308& 6.6587 & 6.372 \\
\hline
\end{tabular}

\section{Implied volatility and small-strike expansions}\label{sec:ImpliedVolatility}
The implied volatility is the Black-Scholes volatility parameter that allows to match observed (or computed) European option prices; 
it obviously depends on strikes and maturities (see for example~\cite{GatheralBook} for more details).
Given a model, the classical route to compute the implied volatility is 
(i) to compute the price of the Call (or Put) option, and (ii) to invert the Black-Scholes formula.
Both steps are numerically demanding, and scarcely provide insights on the behaviour
of the implied volatility smile.
Another route, which has motivated the use of asymptotic methods in finance, 
is to obtain closed-form expansions for the smile (for small/large maturities, or strikes);
however, being asymptotic results, they may lose accuracy when some parameters are not small/large enough.
The `classical' approximation by Hagan et al~\cite{HLW} is such a formula, 
which has been used extensively by practitioners, despite exhibiting flaws--namely arbitrage--in some regions.
This anomaly can in principle be fixed if one accounts for the accumulation of mass at zero due to the Dirichlet boundary condition. 
Let us recall a few (model-independent) results regarding small-strike asymptotics of the implied volatility.
For any strike $K>0$ and maturity $T>0$, let us denote by~$I_T(K)$ the implied volatility.
In the presence of strictly positive mass at zero, the small-strike tail of the implied volatility 
satisfies~\cite{RogerLee}:
\begin{equation}\label{BoundSlope}
\limsup_{K\downarrow 0}\frac{I_T(K)}{\sqrt{|\log K|}}=\sqrt{\frac{2}{T}}. 
\end{equation}
This behaviour was recently refined by De Marco, Hillairet and Jacquier~\cite{DMHJ}, 
and later by Gulisashvili~\cite{GulisashviliMass}.
Assuming that $\PP(X_T\leq K)-\PP(X_T=0)=\mathcal{O}((|\log K|^{-3/2})$ as~$K$ tends to zero,
De Marco, Hillairet and Jacquier~\cite[Proposition 3.1]{DMHJ} 
derive the small-strike asymptotic formula
\begin{equation}\label{DHMJformula}
I_T(K) = \sqrt{\frac{2|\log K|}{T}} + \frac{\mathcal{N}^{-1}(\mathrm{m}_T)}{\sqrt{T}}
 + \frac{(\mathcal{N}^{-1}(\mathrm{m}_T))^2+2}{2\sqrt{2T|\log K|}}
 + \frac{(\mathcal{N}^{-1}(\mathrm{m}_T))}{4|\log K|\sqrt{T}}
 + \mathcal{O}\left(|\log K|^{-3/2}\right),
\end{equation}
where $\mathrm{m}_T := \PP(X_T=0)$ is the mass at the origin, 
and $\mathcal{N}$ the Gaussian cumulative distribution function
(an alternative formulation of~\eqref{DHMJformula} canbe found in~\cite{GulisashviliMass}).

\subsection{Comparison with Ob{\l}{\'o}j~\cite{Obloj}}
\begin{figure}[htb!]
\begin{center}
\includegraphics[scale=0.4]{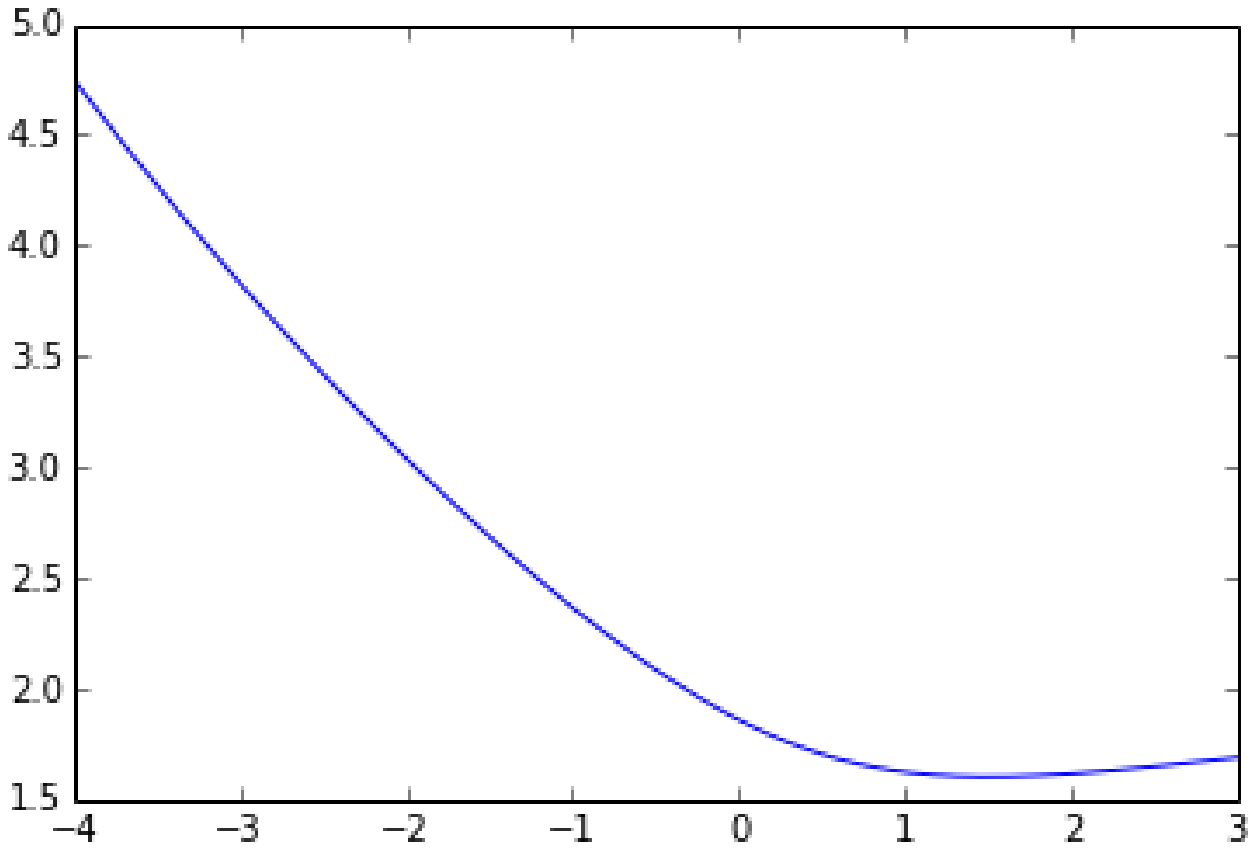}
\includegraphics[scale=0.4]{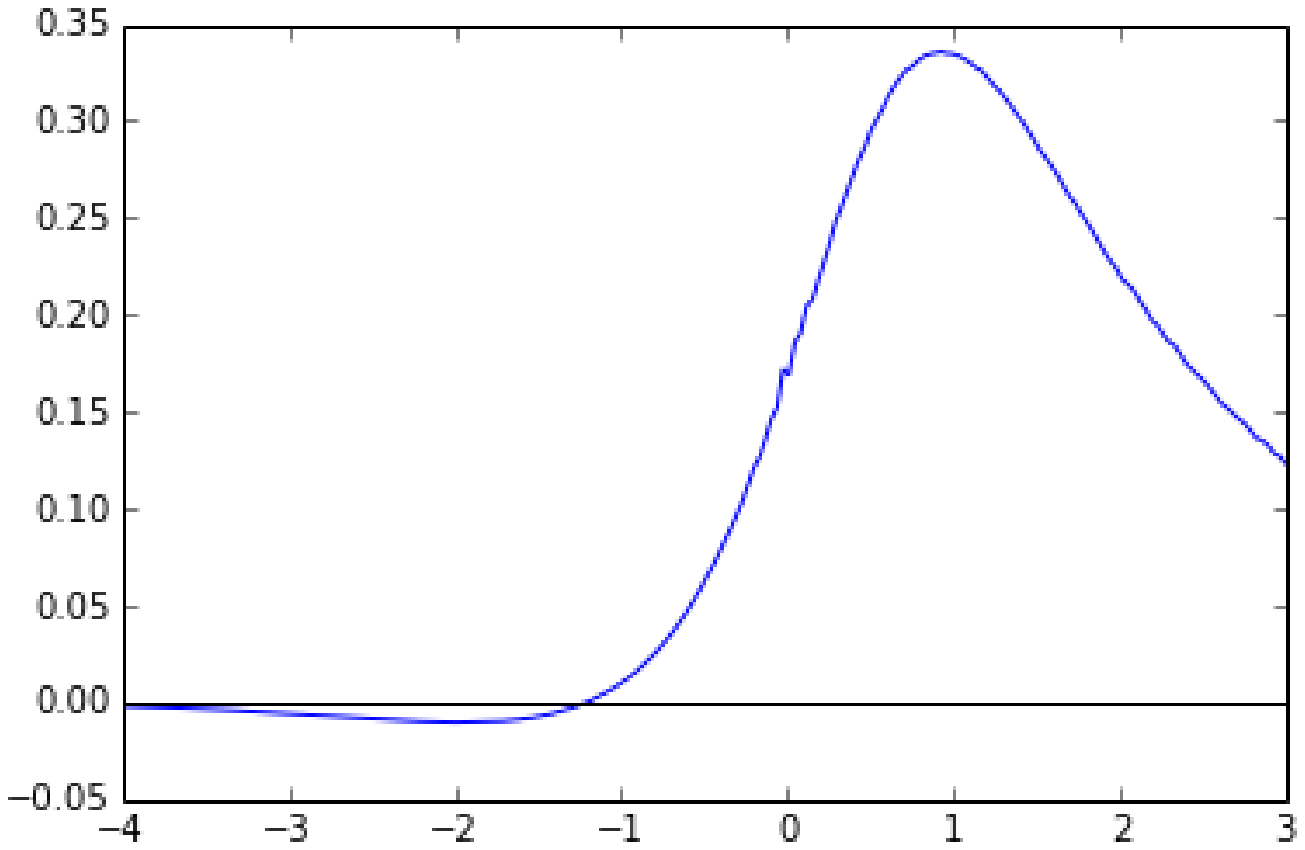}
\caption{Density (right) of the log process~$\log(X)$ 
obtained from the implied volatility expansion~\cite{Obloj} (left) with 
$(\nu, \beta, \rho, x_0, y_0, T) = (0, 1, 0.6, 0.05, 0.5, 1.2)$.
The mass at zero, computed using~\eqref{eq:MassAtZeroCalc} is equal to $4.5\%$.}
\end{center}
\label{fig:OblojDensity}
\end{figure}

\begin{figure}[htb!]
\begin{center}
\includegraphics[scale=0.5]{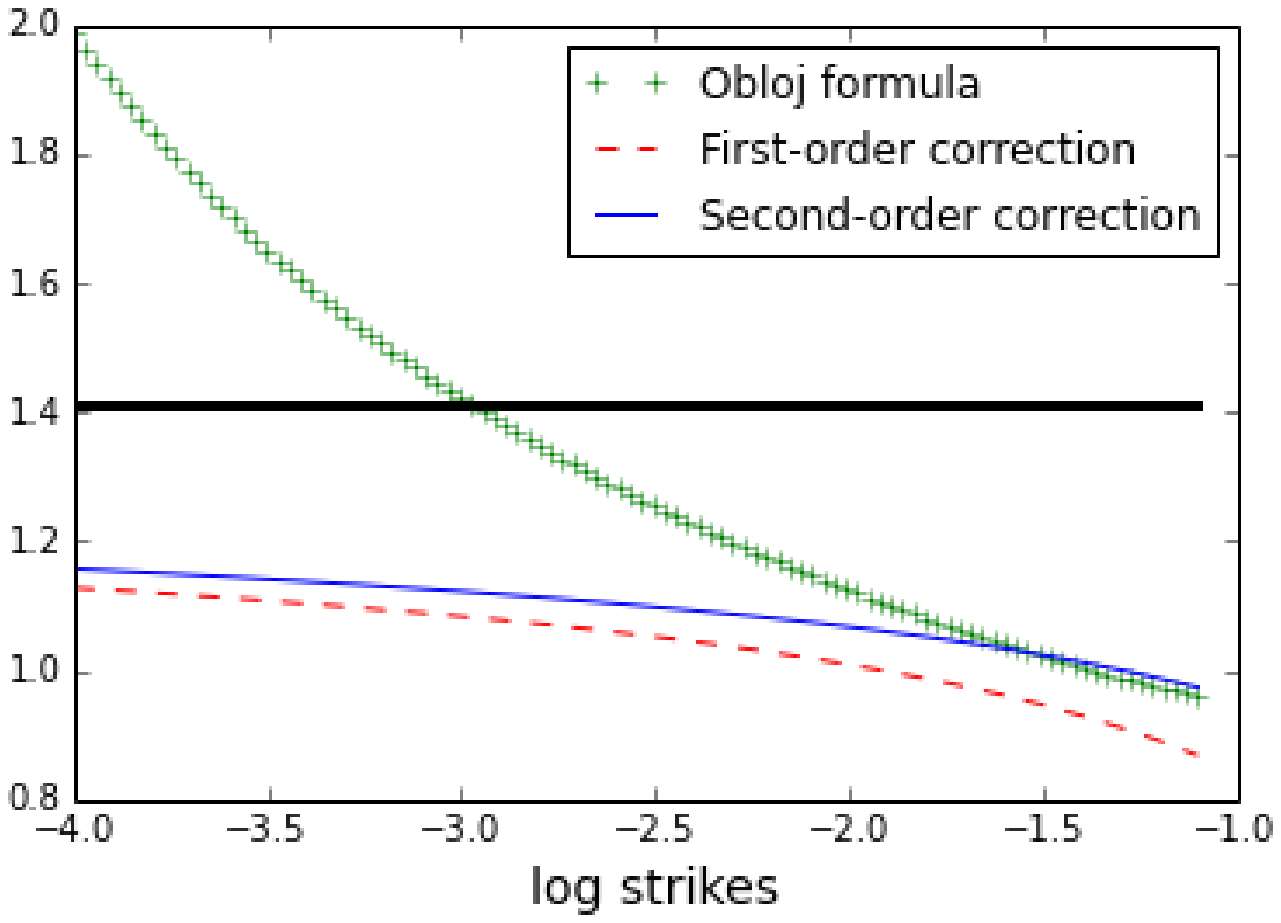}
\includegraphics[scale=0.5]{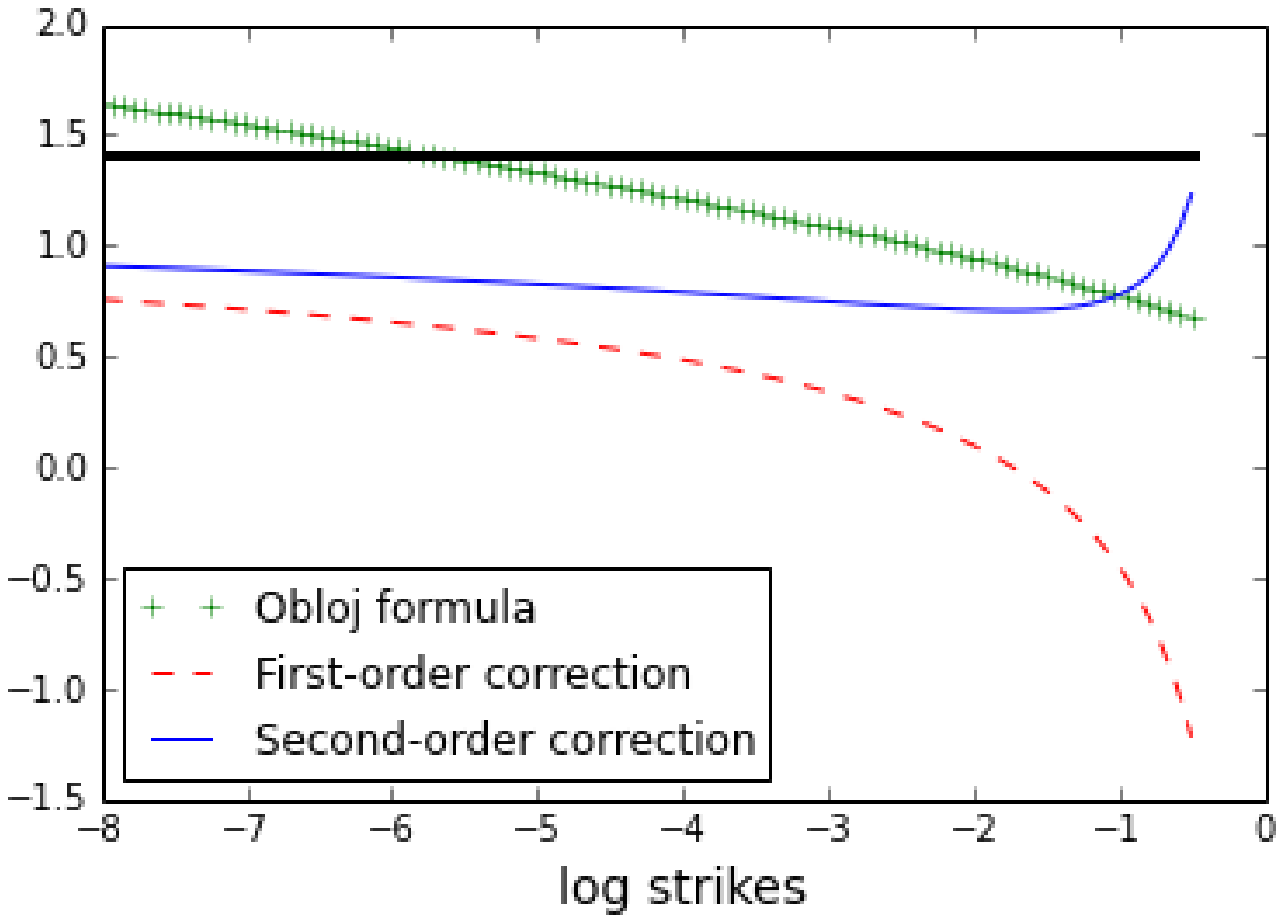}
\caption{
The black line marks the level $\sqrt{2}$. 
The parameters are
$(\nu, \beta, \rho, x_0, y_0, T) = (0.3, 0, 0, 0.35, 0.05, 10)$ for the left plot, 
and 
$(\nu, \beta, \rho, x_0, y_0, T) = (0.6, 0.6, 0, 0.08, 0.015, 10)$
for the right graph. 
Ob{\l}{\'o}j's implied volatility expansion clearly violates this upper bound in both cases.
The large-time mass is equal to $28.3\%$ for the left plot and $3.1\%$ for the right one.}
\end{center}
\label{fig:OblojVolTails}
\end{figure}

Ob{\l}{\'o}j~\cite{Obloj} derived an--refined version of the one in~\cite{HLW}--implied volatility expansion in the SABR model.
However, as we illustrate in Figure~\ref{fig:OblojDensity}, 
this formula exhibits arbitrage for small strikes. 
As explained by Gatheral~\cite[Proof of Lemma 2.2]{Gatheral}, the density of the log price~$\log(X)$ (or log forward rate)
can be expressed directly in terms of the implied volatility,
and negative densities obviously yield arbitrage opportunities.
In Figure~\ref{fig:OblojVolTails}, we visually quantify how `wrong' Hagan's expansion is 
for small strikes in the presence of a mass at the origin.
We plot $k\mapsto I_T(\E^{k})\sqrt{T/|k|}$, which,
from~\eqref{DHMJformula} has to be bounded by $\sqrt{2}$ in order to avoid arbitrage,
and compare it to the first and second order of~\eqref{DHMJformula}
using~\eqref{eq:Mass0LargeTime} to compute the (large-time) mass at zero.
We consider two parameter sets, one for which the large-time mass is small,
and one for yielding a large mass at the origin.
As the mass becomes small, Hagan's (or Ob{\l}{\'o}j's) approximation becomes more accurate.
This holds in particular as the parameter~$\beta$ gets close to one, 
as indicated in Section~\ref{sec:LargetimeuncorNum} above.
In the limit as $\beta=1$, the mass becomes null.

\subsection{Comparison with Antonov-Konikov-Spector~\cite{AntonovWings}}
In the uncorrelated case $\rho=0$, Antonov, Konikov and Spector~\cite{AntonovWings}
derived the double integral formula for the price of a Call option:
$$
\mathbb{E}(X_T-K)_+ = (X_0-K)_+ + \frac{2\sqrt{X_0 K}}{\pi}\left\{\int_{s_-}^{s_+}\frac{\sin(\eta \varphi(s))}{\sinh(s)}G(\nu^2 T, s)\D s
+ \sin(\eta \pi)\int_{s_+}^{\infty}\frac{\exp(-\eta \psi(s))}{\sinh(s)}G(\nu^2 T, s)\D s
\right\},
$$
where
$\eta:=1/|2(\beta-1|$, $q:=\frac{K^{1-\beta}}{1-\beta}$, $q_0:=\frac{X_0^{1-\beta}}{1-\beta}$, 
$s_{\pm}:=\mathrm{arcsinh}\left(\frac{\nu}{y_0}|q\pm q_0|\right)$, 
$$
\varphi(s) := 2\arctan\sqrt{\frac{\sinh(s)^2 - \sinh(s_-)^2}{\sinh(s_+)^2 - \sinh(s)^2}}
\qquad\text{and}\qquad
\psi(s) := 2\mathrm{arctanh}\sqrt{\frac{\sinh(s)^2 - \sinh(s_+)^2}{\sinh(s)^2 - \sinh(s_-)^2}}.
$$
The function $G$ is defined as the integral
$$
G(t,s) := \frac{2\exp\left(-t/8\right)}{t^{3/2}\sqrt{\pi}}
\int_{s}^{\infty}u\sqrt{\cosh(u)-\cosh(s)}\exp\left(-\frac{u^2}{2t}\right)\D u.
$$

\begin{figure}[htb!]
\begin{center}
\includegraphics[scale=0.4]{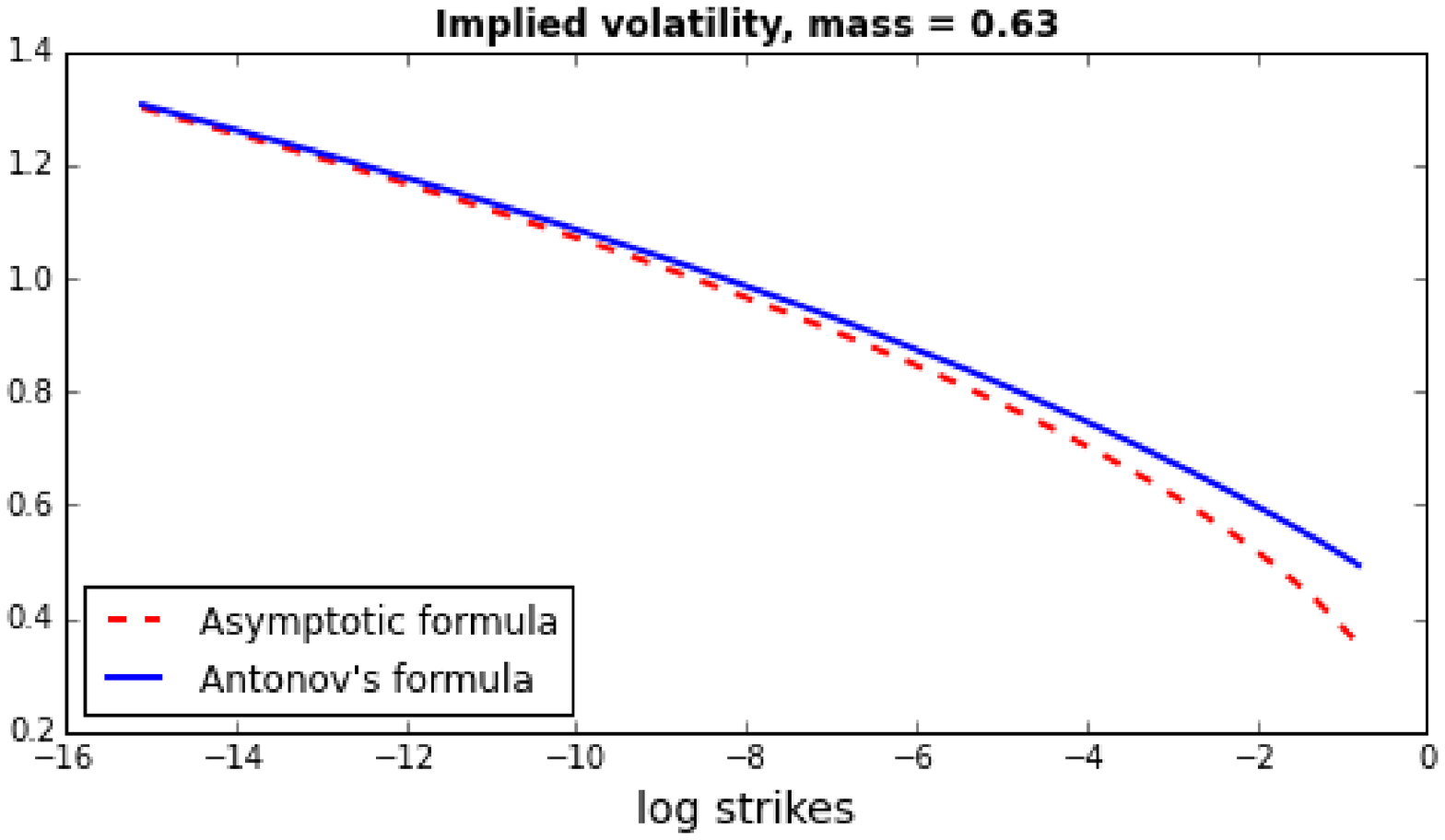}
\end{center}
\label{fig:Antonov}
\end{figure}

In Figure~\ref{fig:Antonov}, we compare the smile
obtained from the Antonov-Konikov-Spector's formula 
(computing the double integral and numerically inverting the Black-Scholes formula) 
and the closed-form tail formula~\eqref{DHMJformula}
using the large-maturity mass at zero computed from~\eqref{eq:Mass0LargeTime}.
Following~\cite{Antonov}, we consider the following set of parameters:
$(\nu, \beta, \rho, x_0, y_0, T) = (0.8, 0.1, 0, 0.1, 0.15, 20)$,
for which the large-maturity mass at zero~\eqref{eq:Mass0LargeTime} is equal to~$63\%$.


\section{Conclusion}
The SABR model is a pillar of mathematical modelling on fixed income desks, 
but suffers from some issues in low interest rate environments, 
where the process can hit the origin with non-zero probability, creating, in most existing approximations
(used by practitioners) arbitrage opportunities.
In this paper, we endeavour to provide accurate estimates for this mass at zero in order 
(i) to quantify the error made by existing approximations, 
and~(ii) to suggest an alternative parameterisation of the implied volatility smile for low strikes,
ensuring arbitrage opportunities do not occur.

\appendix
\section{Proofs of Section~\ref{sec:MassZero}}

\subsection{Proof of Proposition~\ref{prop:expansionMy}}\label{sec:ProofPropExpansion}
Our proof is inspired by~\cite{Gerhold}, which is based on an inverse Laplace transform approach.
From~\cite[Page 645]{Borodin}, the Laplace transform of the function~$m_y$ has a closed-form representation, namely,
whenever $\mu>-3/2$ and $z>0$,
$$
m_y(\mu,z) = \mathcal{L}_u^{-1}\left(\frac{\Gamma(\mu+\frac{1}{2}+\sqrt{u})}{\Gamma(1+2\sqrt{u})}\mathfrak{M}_{-\mu,\sqrt{u}}(2z)\right),
$$
where the function $\mathfrak{M}$ is related to the Kummer function $\M$ function via the identity
$$
\mathfrak{M}_{n,m}(x) \equiv x^{m+1/2}\exp\left(-\frac{x}{2}\right)\M\left(m-n+\frac{1}{2}, 2m+1, x\right).
$$
Therefore, we can write, for some $R\in\mathbb{R}$,
\begin{equation}\label{m:Gerhold}
m_{y}(\mu,z) = \frac{\E^{-z}}{2\I \pi}\int_{R-\I\infty}^{R+\I\infty}\E^{uy}\frac{\Gamma(\mu+\frac{1}{2}+\sqrt{u})}{\Gamma(1+2\sqrt{u})}(2z)^{\frac{1}{2}+\sqrt{u}}
\M\left(\mu+\frac{1}{2}+\sqrt{u}, 1+2\sqrt{u}, 2z\right)\D u.
\end{equation}
Since we wish to determine the behaviour of $m_y$ as $y$ (equivalently, $t$) tends to zero, we need to understand the limit of the integrand as $u$ tends to infinity.
The following asymptotic relations hold uniformly in $v$, as $v=\sqrt{u}$ tends to infinity:
\begin{equation}\label{eq:AsymptoticGammaM}
\Gamma(1+v) = \sqrt{2\pi} \E^{-v} v^{v+1/2}\left[1+\mathcal{O}(v^{-1})\right]
\qquad\text{and}\qquad
\M\left(\mu+\frac{1}{2}+v, 1+2v, 2z\right) \sim \E^{z}.
\end{equation}
The first one is standard~\cite[Section 3.5]{Miller}.
As for the second one, the representation~\eqref{Kummer} yields 
$$
\M\left(\frac{1}{2}+v+\mu, 1+2v, 2z\right) =
\sum_{k=0}^{\infty} \gamma_k\frac{(2z)^k}{k!},
$$
where, for $k\geq 0$,
$$
\gamma_k := \frac{(\mu+v+\frac{1}{2})\cdots(\mu+v+k-\frac{1}{2})}{(1+2v)(2+2v)\cdots(k+2v)}.
$$
Clearly $|\gamma_k|\leq 2^{-k}$ and $\gamma_k \sim 2^{-k}$ as~$v$ tends to infinity, 
and from~\cite[Formula 13.6.3]{Abramowitz}, we have
$$
\M\left(\frac{1}{2}+v+\mu, 1+2v, 2z\right) \leq \M\left(\frac{1}{2}+v, 1+2v, 2z\right)
=\Gamma(1+v)\E^{z}\left(\frac{1}{2}z\right)^{-v}\mathrm{I}_{v}(z), 
$$
where again $\mathrm{I}_{v}$ denotes the modified Bessel function of the first kind~\cite[Page 638]{Borodin}, 
so that, using~\eqref{eq:AsymptoticGammaM} and~\cite[Equation~9]{Gerhold}, 
we have, uniformly in $v$,
$$
\left| \M\left(\frac{1}{2}+v+\mu, 1+2v, 2z\right)\right|
\leq \Gamma(1+v)\E^{z}\left(\frac{z}{2}\right)^{-v}\mathrm{I}_{v}(z)
 = \E^{z}\left(1 + \mathcal{O}\left(v^{-1}\right)\right).
$$

Therefore the integrand in~\eqref{m:Gerhold} reads, as $u$ tends to infinity, 
\begin{align*}
\Phi(u, y, z)
 &  \equiv \E^{uy}\frac{\Gamma(\mu+\frac{1}{2}+\sqrt{u})}{\Gamma(1+2\sqrt{u})}(2z)^{\frac{1}{2}+\sqrt{u}}
\M\left(\mu+\frac{1}{2}+\sqrt{u}, 1+2\sqrt{u}, 2z\right)\\
 & \sim \E^{v^2 y + v + z}z^{v+\frac{1}{2}} v^{\mu-v-\frac{1}{2}} 2^{-v}\\
 & = \exp\left[v^2 y + \alpha v + \left(\mu-\frac{1}{2}-v\right)\log(v) + z + \frac{1}{2}\log(z)\right]
 =: \exp\left(\psi_y(u) + z + \frac{1}{2}\log(z)\right).
\end{align*}
where $\alpha:=1 + \log(z) - \log(2)\in\mathbb{R}$, and where the function $\psi_y$ is defined by
\begin{equation}\label{m:Gerhold2}
\psi_y(u)
 \equiv u y-\frac{1}{2}\sqrt{u}\log(u) + \alpha\sqrt{u} + \frac{1}{2}\left(\mu - \frac{1}{2}\right)\log(u).
\end{equation}
For $y>0$ small enough, the saddlepoint equation $\partial_u \psi_y(u) = 0$, or
$$
2\mu - 1 + 4 u y + 2(\alpha-1)\sqrt{u} -\sqrt{u}\log(u) = 0,
$$
(namely~\eqref{eq:Saddlepoint}) admits a solution
$u_y>0$.
This saddlepoint equation can be rewritten as
\begin{equation}\label{eq:saddlepointEq}
y = \frac{\log (u_y)}{4 \sqrt{u_y}}+\frac{1-\alpha}{2\sqrt{u_y}} - \frac{\left(\mu -1/2 \right)}{2 u_y}.
\end{equation}
\begin{remark}
Note that the saddlepoint equation also reads
$$
y = \frac{\log(u_0)}{2 \sqrt{2u_0}} - \frac{\rho}{\sqrt{2 u_0}}-\frac{4 \mu -2}{4u_0},
$$
where $\rho  := \log(z/\sqrt{2})$ and $u_0:=2u_y$, which is reminiscent of that of~\cite{Gerhold}.
In fact, the saddlepoint equation above does not admit a unique solution;
in order for the latter to be continuous (as a function of $y$), one should take the largest solution.
\end{remark}
Following~\cite{Gerhold}, we deform the integration contour in~\eqref{m:Gerhold} 
around the saddlepoint~$u_y$ to obtain
\begin{equation}\label{eq:Integral_my}
m_{y}(\mu,z) = \frac{\E^{-z}}{2\I \pi}\int_{R-\I\infty}^{R+\I\infty}\Phi(u, y, z)\D u
\sim \frac{\sqrt{z}}{2\I \pi}\int_{R-\I\infty}^{R+\I\infty}\E^{\psi_y(u)}\D u
\sim \frac{\sqrt{z}}{2\I \pi}\int_{u_y-\I\infty}^{u_y+\I\infty}\E^{\psi_y(u)}\D u,
\end{equation}
as $y$ tends to zero.
Let $\lambda$ denote the real integration variable, so that
$u = u_y + \I \lambda$.
Around the saddlepoint ($\lambda = 0$), we have the uniform Taylor series expansions:
\begin{align*}
\sqrt{u} & = \sqrt{u_y} + \frac{\I \lambda}{2\sqrt{u_y}} + \frac{\lambda^2}{8u_y^{3/2}}
+ \mathcal{O}\left(\frac{\lambda^3}{u_y^{5/2}}\right),
\qquad\qquad
\log{u} = \log{u_y} + \frac{\I \lambda}{u_y} + \frac{\lambda^2}{2u_y^{2}}
+ \mathcal{O}\left(\frac{\lambda^3}{u_y^{3}}\right),\\
\sqrt{u}\log{u} & = \sqrt{u_y}\log{u_y} + \frac{(2 +\log(u_y))\I \lambda}{2\sqrt{u_y}}
 + \frac{\log(u_y) \lambda^2}{8 u_y^{3/2}}
+ \mathcal{O}\left(\frac{(1+\log(u_y))\lambda^3}{u_y^{5/2}}\right),
\end{align*}
so that
\begin{equation}\label{eq:PsiExpansion}
\psi_y(u) = u_y y + \sqrt{u_y}\left(\alpha-\frac{\log(u_y)}{2}\right) - \frac{\log(u_y)}{4}
 + \frac{\mu\log(u_y)}{2}
 - M_y \lambda^2  + \mathcal{O}\left[\frac{\lambda^3(1+\log(u_y))}{u_y^{5/2}}\right]
\end{equation}
where the coefficients in front of~$\lambda$ cancel out from the saddlepoint equation~\eqref{eq:Saddlepoint}, 
and where
$$
M_y := \frac{\log(u_y)}{16 u_y^{3/2}} - \frac{\alpha }{8 u_y^{3/2}} + \frac{1-2\mu}{8u_y^2},
$$
as defined in Proposition~\ref{prop:expansionMy}.
By bootstrapping (see Section~\ref{sec:ExpansionM} for details), the expansion 
\begin{equation}\label{eq:BootstrapUy}
u_y = \frac{\log(y)^2}{4y^2}\left[1- \frac{2\log\log(1/y)}{\log(y)} + 
 \frac{\log(z^2)}{\log(y)} + o\left(\frac{1}{\log(y)}\right)\right]
\end{equation}
holds
for the saddlepoint as $y$ tends to zero, and implies
\begin{equation}\label{eq:bootstrapM}
M_y = \frac{y^3}{\log(y)^2}\left[1+\mathcal{O}\left(\frac{\log|\log(y)|}{\log(y)}\right)\right].
\end{equation}
Since~\eqref{eq:Integral_my} can be rewritten as 
\begin{align*}
m_{y}(\mu,z)
& \sim \frac{\sqrt{z}}{2\I \pi}\int_{u_y-\I h}^{u_y+\I h}\E^{\psi_y(u)}\D u\\
& \sim \frac{\sqrt{z}}{2\pi}
\exp\left[u_y y  + \sqrt{u_y}\left(\alpha-\frac{\log(u_y)}{2}\right) - \frac{\log(u_y)}{4}
 + \frac{\mu\log(u_y)}{2}\right]
\int_{-h}^{h} \E^{-M_y \lambda^2}\D \lambda,
\end{align*}
we need to determine an estimate for the last integral on the compact interval~$[-h,h]$.
As explained below--where the tail integrals are taken into account--the choice $h:=\log(y)^2 / y^{3/2}$
is in fact the right one, and it is then easy to show that, as~$y$ tends to zero,
$$
\int_{-h}^{h}\E^{-M_y \lambda^2}\D \lambda 
  = \frac{\sqrt{\pi}|\log(y)|}{y^{3/2}} + \Oo\left(\E^{-\log(y)^2}y^{-3/2}\right),
$$
which then implies
\begin{align}
m_{y}(\mu,z)
& \sim \frac{\sqrt{z}}{2\pi}
\exp\left[u_y y + \sqrt{u_y}\left(\alpha-\frac{\log(u_y)}{2}\right) - \frac{\log(u_y)}{4}
 + \frac{\mu\log(u_y)}{2}\right]
\frac{\sqrt{\pi}|\log(y)|}{y^{3/2}}\nonumber\\
 & = \frac{\sqrt{z}}{2\pi} \exp\left[ \left(\frac{1}{2}-\mu\right)+ \frac{\log(u_y)}{2}\left(\mu - \frac{1}{2}\right) - u_y y + \sqrt{u_y} \right]
\frac{\sqrt{\pi}|\log(y)|}{y^{3/2}}\nonumber\\
 & = \frac{\sqrt{z}}{2\sqrt{\pi}}\exp\left(\frac{1}{2}-\mu\right)
 \frac{|\log(y)|}{y^{3/2}}
u_y^{\frac{1}{2}\left(\mu - \frac{1}{2}\right)}
 \exp\left(- u_y y + \sqrt{u_y}\right)\nonumber\\
& = \frac{z^{\frac{1}{2}}|\log(y)|}{2\sqrt{\pi}}
\exp\left\{-\frac{\log(y)^2}{4y} + \frac{1}{2}\frac{|\log(y)|}{y}
+\left(\frac{1}{2}-\mu\right)\left[1-\frac{1}{2}\log\left(\frac{\log(y)^2}{4y^2}\right)\right]\right\}
\left[\frac{1}{y^{\frac{3}{2}}}+\mathcal{O}\left(y^{\frac{3}{2}}\right)\right],\label{expansionm}
\end{align}
where we used the saddlepoint equation~\eqref{eq:saddlepointEq} in the fourth line.

It now remains to prove that one can indeed neglect the tails of the integration domain, 
where $\Im(u) = \lambda\geq h$.
The analysis of this is similar to that of~\cite[Section 3]{Gerhold}, and we only outline here the main arguments.
First, specify a choice $h:=\log(y)^2 / y^{3/2}$ of integration bounds 
accounting for the main contribution to the integral 
$\int_{u_y-\I\infty}^{u_y+\I\infty}\exp(\psi_y(u))\D u$,
with $\psi_y$ defined in~\eqref{m:Gerhold} 
and where $u_y$ denotes the saddlepoint in~\eqref{eq:saddlepointEq}.
By symmetry, it is clearly sufficient to consider only one side of the tails,
and we shall therefore focus on the positive one
$\int_{u_y +\I h}^{u_y+\I\infty}\E^{\psi_y(u)}\D u$.
The analysis is then split into studying the inner tail $h\leq \lambda <\exp(\log(1/t)^2/4)$ 
and the outer tail $ \lambda \geq \exp(\log(1/t)^2/4)$.
Similarly to~\cite[Equation (10)]{Gerhold}, the estimate
\begin{align*}
\int_{u_y +\I h}^{u_y+\I\infty}\E^{\psi_y(u)}\D u 
\sim 2 \exp\left\{u_y t+ \frac{1}{8}\log(y)^2 - \exp\left(\frac{\log(y)^2}{8}\right)\right\}
\end{align*}
prevails for the outer tail. 
For any real number~$B$, \cite[Lemma 1]{Gerhold} remains valid for the behaviour of the real part of 
$\sqrt{u}\log(u) +B \sqrt{u}$ with respect to $|\Im(u)|$,
which allows to bound above the inner tail by the value of the integrand at $\lambda=h$
of $-M_y \lambda^2\vert_{\lambda=h}\sim -\frac{1}{2}\log(y)^2$ 
multiplied by the length of the integration path, which is of order $\E^{\log(1/t)^2/4}$;
the relative error is therefore of order $\exp(-\frac{1}{4}\log(y)^2+o(\log(y)^2))$.

The final part of the error analysis in the expansion~\eqref{expansionm} follows
from analogous estimates to~\cite[Table 1]{Gerhold},
and the total (both tails) error resulting from the completion to Gaussian integral
$$
\frac{2}{\sqrt{2M_y}}\int_{h\sqrt{2M_y}}^{\infty}\exp\left(-\frac{1}{2}\omega^2\right)\D \omega 
 \sim \frac{2}{\sqrt{2M_y}}\frac{\exp\left(-\frac{1}{2}\omega^2\right)}{\omega}\Bigg|_{\omega=h\sqrt{M_y}}
 = \exp\left(-\frac{1}{2}\log(t)^2+o(\log(t))\right).
$$
The error $\mathcal{O}(\lambda^3 / u_y^{5/2})$ from the local expansion~\eqref{eq:PsiExpansion} for~$\psi_y$ is of order
\begin{equation}\label{relerlocal}
\mathcal{O}\left(\log(y)^2 \sqrt{y}\right), 
\end{equation}
which is immediate from bootstrapping~\eqref{eq:bootstrapM},~\eqref{eq:BootstrapUy} 
for~$M_y$ and~$u_y$, and from the choice of~$h$,
\begin{align*}
\frac{\lambda^3}{u_y^{5/2}}\leq C \frac{\log(u_y)}{u_y^{3/2}} \frac{1}{u_y} \frac{\log(y)^2}{y^{3/2}}\sim C \log(y)^2 \sqrt{y}.
\end{align*}
Hence the total relative error is dominated by the error~\eqref{relerlocal} from the local expansion if~$M_y$ is not expanded, and by the relative error~\eqref{eq:bootstrapM} of~$M_y$ 
if one consider its bootstrapping expansion.

\subsubsection{Justification of the expansion for $M_y$}\label{sec:ExpansionM}
We define
$$
\widetilde{M}_y:=\frac{\log(u_y)}{16 u_y^{3/2}}-\frac{\alpha}{8 u_y^{3/2}}.
$$
The term $(1-2\mu)/(8u_y^2)$ in the definition~\eqref{eq:My} of~$M_y$ is of higher order,
so that  we can work with the simpler expression~$\widetilde{M}_y$ 
in the bootstrapping expansion and the error analysis instead of~$M_y$.
With $\alpha=\rho+1-\frac{1}{2}\log(2)$ and $\widetilde u\equiv u_y /2$, 
the approximation of the saddlepoint simplifies to
$$
M_y=\frac{\log(u_y)}{16 u_y^{3/2}}-\frac{\alpha}{8 u_y^{3/2}}
 = \frac{\log(u_y)}{16 u_y^{3/2}}-\frac{\rho+1}{8 u_y^{3/2}}+\frac{\log(2)}{16 u_y^{3/2}}
 = \frac{1}{4}\left(\frac{\sqrt{2}\log(\widetilde u)}{16  \widetilde u^{3/2}}-\frac{\sqrt{2}(\rho+1-\log(2))}{8 \widetilde u^{3/2}}\right).
$$
Thus~$M_y$ is up to constants of the same form as~\cite[Equation (12)]{Gerhold}.
By bootstrapping, 
$$
M_y \sim \frac{y^3}{\log(y)^2}\left[1+\mathcal{O}\left(\frac{\log|\log(y)|}{\log(y)}\right)\right].
$$
Indeed,  the saddlepoint equation~\eqref{eq:Saddlepoint}
\begin{align*}
y=\frac{\log(2 u_y)}{2 \sqrt{2 (2u_y)}}-\frac{\rho}{\sqrt{2 (2u_y)}}-\frac{4 \mu -2}{4 (2u_y)},
\end{align*}
when setting $c(u)\equiv \left(\frac{\log(\sqrt{u})}{\sqrt{2}}-\frac{\rho}{\sqrt{2}}+\frac{k}{4\sqrt{u}}\right)$, 
$\rho=\log\left(\frac{z}{\sqrt{2}}\right)$, $k:=4 \mu -2$ and $u_0 = 2 u_y$ 
becomes
$\sqrt{u_0}\equiv y^{-1}c(u_0)$,
where
$\log(\sqrt{u_0}) = \log\left(c\left(u_0\right)-\log(y)\right)$.
Hence, bootstrapping as in \cite{Gerhold} yields
\begin{align*}
u_0&=\frac{1}{y^2}\left(\frac{\log\left(1/y\right)}{\sqrt{2}} +\frac{\log(c\left(u_0\right))}{\sqrt{2}}-\frac{\rho}{\sqrt{2}}+\frac{k}{4\sqrt{u_0}}\right)^2\\ 
&=\frac{1}{y^2}\left(\left(\frac{\log\left(1/y\right)}{\sqrt{2}}\right)^2+ 2\left(\frac{\log\left(1/y\right)}{\sqrt{2}}\right)\left(\frac{\log(c\left(u_0\right))}{\sqrt{2}}-\frac{\rho}{\sqrt{2}}+\frac{k}{4\sqrt{u_0}}\right) +\left(\frac{\log(c\left(u_0\right))}{\sqrt{2}}-\frac{\rho}{\sqrt{2}}+\frac{k}{4\sqrt{u_0}}\right)^2\right)\\
&=\frac{\left(\log\left(1/y\right)\right)^2}{2y^2}
\left[1+ \left(\frac{2 \sqrt{2}}{\log\left(1/y\right)}\right)\left(\frac{\log(c\left(u_0\right))}{\sqrt{2}}-\frac{\rho}{\sqrt{2}}+\frac{k}{4\sqrt{u_0}}\right) + \frac{2}{\left(\log\left(1/y\right)\right)^2}\left(\frac{\log(c\left(u_0\right))}{\sqrt{2}}-\frac{\rho}{\sqrt{2}}+\frac{k}{4\sqrt{u_0}}\right)^2\right].
\end{align*}
Now expanding around $\log(1/y)$,
\begin{align*}
 \frac{\log\left(c(u_0)\right)}{\sqrt{2}}
\sim \frac{\log\left(\log(1/y)\right)}{\sqrt{2}}-\frac{\log(2)}{2\sqrt{2}}+\frac{\log\left(c(u_0)-\rho + \frac{k}{2 \sqrt{2u_0}}\right)}{\sqrt{2} \log(1/y)},
\end{align*}
and using the fact that both
$$
 -\frac{2\sqrt{2}}{\log(y)}
\left(\frac{k}{4 \sqrt{u}} - \frac{\log\left(c(u)-\rho+\frac{k}{2 \sqrt{2u_0}}\right)}{\sqrt{2}\log(y)}\right)
\qquad\text{and}\qquad
\frac{2}{\log(y)^2}
\left(\frac{\log\left(c(u_0)\right)-\rho}{\sqrt{2}}+\frac{k}{4\sqrt{u_0}}\right)^2
$$
are of order $o\left(1/\log(y)\right)$, we obtain, by collecting terms,
\begin{align*}
2u_y= \frac{\log(y)^2}{2y^2}
\Bigg(1 - \frac{2 \log(-\log(y))}{\log(y)} + \frac{2 \rho + \log(2)}{\log(y)}+ o\left(\frac{1}{\log(y)}\right)
\Bigg).
\end{align*}
Similarly,
$$
u_0^{3/2}
 = \frac{1}{y^3}\left[\frac{-\log(y)}{\sqrt{2}} +\frac{\log(c(u))}{\sqrt{2}}-\frac{\rho}{\sqrt{2}}+\frac{k}{4\sqrt{u}}\right]^3
\sim
\frac{-\log(y)^3}{2\sqrt{2}y^3}
\left[1 - \frac{\log(-\log(y))}{\log(y)} + \frac{2 \rho+\log(2)}{2\log(y)} + o\left(\frac{1}{\log(y)}\right)
\right]^3,
$$
hence $u_y^{3/2}\sim \left(\log\left(1/y\right)\right)^2 / (8y^2)$; further,
$$
\log(u_0)
 = -2\left(\log(y) - \log(c\left(u\right)) \right)
 \sim -2\log(y) +2\log(-\log(y))-\log(2) - \frac{2\log\left(c(u)-\rho + \frac{k}{2 \sqrt{2u}}\right)}{\log(y)},
$$
so that, by bootstrapping we also recover the form of~\cite[Equation (13)]{Gerhold}, 
at $\widetilde{u} = \frac{1}{2}u_y$:
$$
M_y
 = \frac{1}{4}\left[\frac{\sqrt{2}\log( \widetilde u)}{16 \widetilde u^{3/2}} 
  - \frac{\sqrt{2}(\rho+1-\log(2))}{8 \widetilde u^{3/2}}\right]
 = \frac{y^3}{2 \log(y)^2} \left[1 + \mathcal{O}\left(\frac{\log(-\log(y))}{\log(y)}\right) \right].
$$


\end{document}